\documentclass[12pt, a4paper]{article}
\pdfoutput=1

\usepackage{epsf}

\usepackage{epsfig,graphics}
\usepackage{graphicx}
\usepackage {epsfig}
\usepackage {subfigure}
\usepackage {tabularx} 
\usepackage{rotate}	
\usepackage{slashed}
\usepackage{color}
\usepackage{amsmath}
\usepackage{amsfonts}
\usepackage{amssymb}
\usepackage{graphicx}
\usepackage{cite}

\def\yzero{\smash{\hbox{$y\kern-4pt\raise1pt\hbox{${}^\circ$}$}}}

\def\b{\beta}

\def\beq{\begin{equation}}
\def\eeq{\end{equation}}
\def\beqa{\begin{eqnarray}}
\def\eeqa{\end{eqnarray}}

\def\-{\hphantom{-}}

\def\s2{\frac{1}{\sqrt2}}

\def\beq{\begin{equation}}
\def\eeq{\end{equation}}
\def\beqa{\begin{eqnarray}}
\def\eeqa{\end{eqnarray}}

\def\IF{\relax{\rm I\kern-.18em F}}
\def\II{\relax{\rm I\kern-.18em I}}
\def\IP{\relax{\rm I\kern-.18em P}}
\def\IC{\relax\hbox{\kern.25em$\inbar\kern-.3em{\rm C}$}}
\def\IR{\relax{\rm I\kern-.18em R}}

\def\Dsl{\,\raise.15ex\hbox{/}\mkern-13.5mu D} 
\def\IZ{Z\kern-.4em  Z}

\catcode`\@=11   
\newdimen\@rotdimen
\newbox\@rotbox  

\def\@vspec#1{\special{ps:#1}}
\def\@rotstart#1{\@vspec{gsave currentpoint currentpoint translate
   #1 neg exch neg exch translate}}
\def\@rotfinish{\@vspec{currentpoint grestore moveto}}
\def\@rotr#1{\@rotdimen=\ht#1\advance\@rotdimen by\dp#1%
   \hbox to\@rotdimen{\hskip\ht#1\vbox to\wd#1{\@rotstart{90 rotate}%
   \box#1\vss}\hss}\@rotfinish}
\def\@rotl#1{\@rotdimen=\ht#1\advance\@rotdimen by\dp#1%
   \hbox to\@rotdimen{\vbox to\wd#1{\vskip\wd#1\@rotstart{270 rotate}%
   \box#1\vss}\hss}\@rotfinish}%
\def\@rotu#1{\@rotdimen=\ht#1\advance\@rotdimen by\dp#1%
   \hbox to\wd#1{\hskip\wd#1\vbox to\@rotdimen{\vskip\@rotdimen
   \@rotstart{-1 dup scale}\box#1\vss}\hss}\@rotfinish}%
\def\@rotf#1{\hbox to\wd#1{\hskip\wd#1\@rotstart{-1 1 scale}%
   \box#1\hss}\@rotfinish}%
\def\rotate{\@ifnextchar[{\@rotate}{\@rotate[l]}}
\def\@rotate[#1]#2{\setbox\@rotbox=\hbox{#2}\@nameuse{@rot#1}\@rotbox}

\catcode`\@=12

\begin{document}

\makeatletter
\@addtoreset{equation}{section}
\makeatother
\renewcommand{\theequation}{\thesection.\arabic{equation}}
\pagestyle{empty}
\vspace{-0.2cm}
\rightline{}
\rightline{}
\vspace{1.2cm}
\begin{center}


\LARGE{The 750 GeV LHC diphoton excess from a baryon number conserving string model} 
\\[13mm]
  \large{ Christos Kokorelis$^{1, 2}$   \\[6mm]}
\small{
 $^{1}$Department of Physics, National Technical University of Athens, 15780, Athens, Greece\\[-0.3em]
$^{2}$Natural Sciences Division, Hellenic Army Academy, Vari, 16673, Athens, Greece\\[8mm]}
\small{\bf Abstract} \\[7mm]
\end{center}
\begin{center}
\begin{minipage}[h]{15.22cm}
We propose an explanation of the LHC data excess resonance of 750 GeV in the diphoton distribution using D-brane models, with
 gauged baryon number, which accommodate the Standard Model together with vector like exotics. We identify the 750 GeV scalar 
as either the sneutrino (${\tilde \nu}_R$) or as an axion. Using  a bottom-up approch, ${\tilde \nu}_R$ is produced via gluon fusion
when scalar (supersymmetric) partners of vector like quarks and leptons are generated by demanding that 
the corresponding intersections                                                                                                                                                                                          
respect N=1 supersymmetry. When we generate the value of  ${\tilde \nu}_R$ at 750 GeV, by varying the complex 
structure of the torus,  the string scale is limited to be in the range 
$ 10^{14} < M_s < 10^{19}$ GeV. 
Also, generating the Higgs scalars by imposing N=1 supersymmetry on intersections 
fixes naturally, to zero, the coupling of the axion to 
SU(2) gauge bosons  $\propto  F_b \wedge F_b$ and in photon-photon fusion also decouples its coupling  from $G^2$ of color SU(3) by 
simultaneously generating the superpartners of $q_L$, $U_R$, $l_L$, $\nu_R$ quarks and leptons.
\end{minipage}
\end{center}
\newpage
\setcounter{page}{1}
\pagestyle{plain}
\renewcommand{\thefootnote}{\arabic{footnote}}
\setcounter{footnote}{0}

\section{Introduction }
\label{sec1}

Run 2 LHC early data from ATLAS and CMS at an energy $\sqrt{s}=13$ TeV using integrated lunimocities of 
3.2 fb$^{-1}$ and 2.6 fb$^{-1}$ show hints of a new resonance in the diphoton distribution of pp collisions 
at an invariant mass of 750 GeV \cite{atl1, cls1}. 
The corresponding excess in the cross section can be estimated to be $\sigma_{pp\rightarrow \gamma \gamma}^{13 \ TeV} \sim 3 - 13 $ 
fb \cite{atl1, cls1}. 
At the Moriod 2016 conference, the ATLAS and CMS collaborations  \cite{skata1},  \cite{skata2},  \cite{skata3} updated their search, increasing 
the statistical 
significance of the excess around $m_{\gamma \gamma} \approx 750$ GeV (up to 3.9σ in ATLAS and 3.4σ in CMS, locally) but do 
not qualitatively change the main implications, still maintaining a hint for an excess at 750 GeV.\newline
 The origin of the new resonance has been attributed to a lot of
different scenarios \cite{Ha}-\cite{Cho}.  In most of the scenarios the 750 GeV resonance is a scalar (or a 
pseudoscalar) that is produced via gluon fusion ans subsequently decays to two photons, via loops of vector-like fermions.   
In the context of string theories a variety of works has dicsussed the diphoton excess \cite{hec}, \cite{cve1}, \cite{anto1}, \cite{anto3}, \cite{ibaax}, \cite{karo},  \cite{fara1}, \cite{anasta}, \cite{abel}, \cite{cve2}, \cite{li1}, \cite{leo}, \cite{fara2}.
In this work, we use the ${\tilde \nu}_R$ as the source of the diphoton excess (DE) in baryon number conserving non-supersymmetric 
D-brane models, using the model of \cite{kokoD6}.  The presence of N=1 supersymmetry in particular intersections, gives birth to the previously massive ${\tilde \nu}_R$
 that now become massless at the string
scale and survives to low energies.
 Other works which used ${\tilde \nu}_R$ to explain DE, use an R-parity violating background in the MSSM \cite{li2}, \cite{alla}.

This paper is organized as follows. In Section 2 we will discuss the basic structure of the non-supersymmetric intersecting D6-brane model considered.  
We discuss the interpretation of the DE in terms of ${\tilde \nu}_R$ in Section 3, especially the appearance of ${\tilde \nu}_R$,  and extra 
fermions (also scalars), in the 
presence of supersymmetry on intersections. In Section 4 we describe the use of the axion 
in the D6-brane models considered, as a alternative possibility to explain the DE. Our models possess all the ingredients to explain the use of 
the string axion to explain the 750 GeV state in terms of gluon or photon fusion, when superpartners of Higgsinos and some SM ferrmions are 
present because of N=1 supersymmetries preserved at particular intersections.
 We conclude in Section 5.

\section{The quiver Standard Model and its embedding on a string construction}  
\label{sec2}


\subsection{The five stack quiver Standard Model }

\begin{figure*}
\begin{center}
\includegraphics[scale=0.8]{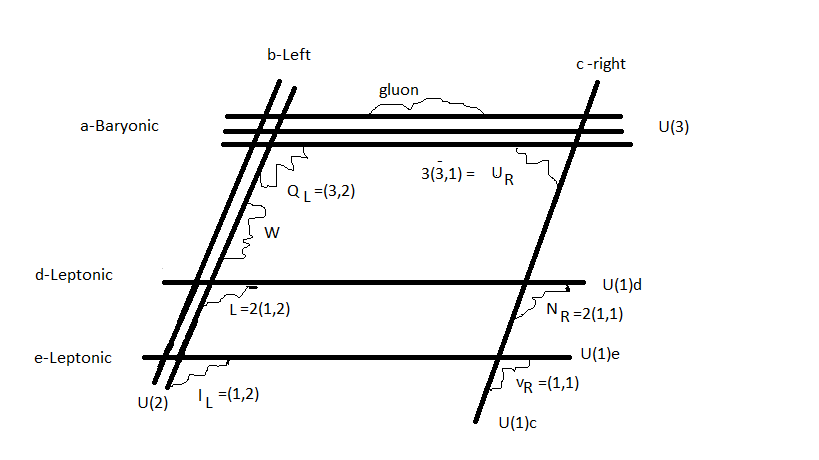}
\end{center}
\caption{\footnotesize Quarks and leptons at the intersecting brane $U(3)_c \times U(2)_w\times U(1)_c \times U(1)_d \times U(1)_e$ model.}
      \label{branes}
\end{figure*}

In intersecting brane constructions chiral fermions appear as open strings streching between brane
 intersecting at angles and gauge bosons living on branes. Each D-brane would give rise to a U(1) and the U(N) gauge group arises from 
N overlapping 
D-branes (stacks).  By considering $a$ 
stacks of D-brane configurations with 
$N_a, a = 1, \cdots, N$, parallel branes one gets the gauge group 
 $U(N_1) \times U(N_2) \times \cdots \times U(N_a)$. 
Each $U(N_i)$ factor will give rise to an $SU(N_i)$
charged under the associated $U(1_i)$ gauge group factor that appears in 
the decomposition $SU(N_a) \times U(1)_a$. In  this paper, we are using the five stack D6-brane string model of \cite{kokoD6}. The initial gauge group  
of the model is $U(3)_c \times U(2)_b 
\times U(1)_c \times U(1)_d \times U(1)_e$ or $SU(3)_c \times SU(2)_w \times U(1)_b \times U(1)_c \times U(1)_d \times U(1)_e$ at the string 
scale. The model
has the interesting properties of elevating the global symmetries of the Standard model (SM) , namely Baryon B and Lepton number L, to 
local gauge symmetries.  
This model may be used 
to explain the diphoton excess observed from ATLAS and CMS collaborations.
The representation content of the Standard Model is seen at figure (\ref{branes})
and table (\ref{spectrum8}) charged under the five U(1) symmetries $Q_a, Q_b, Q_c, Q_d, Q_e$.

\begin{table}[htb] \footnotesize
\renewcommand{\arraystretch}{1.2}
\begin{center}
\begin{tabular}{|c|c|c|c|c|c|c|c|c|}
\hline
Matter Fields & & Intersection & $Q_a$ & $Q_b$ & $Q_c$ & $Q_d$ & $Q_e$& Y
\\\hline
 $Q_L$ &  $(3, 2)$ & $I_{ab}=1$ & $1$ & $-1$ & $0$ & $0$ & $0$& $1/6$ \\\hline
 $q_L$  &  $2(3, 2)$ & $I_{a b^{\ast}}=2$ &  
$1$ & $1$ & $0$ & $0$  & $0$ & $1/6$  \\\hline
 $U_R$ & $3({\bar 3}, 1)$ & $I_{ac} = -3$ & 
$-1$ & $0$ & $1$ & $0$ & $0$ & $-2/3$ \\\hline    
 $D_R$ &   $3({\bar 3}, 1)$  &  $I_{a c^{\ast}} = -3$ &  
$-1$ & $0$  & $-1$ & $0$ & $0$ & $1/3$ \\\hline    
$L$ &   $2(1, 2)$  &  $I_{bd} = -2$ &  
$0$ & $-1$ & $0$ & $1$ & $0$ & $-1/2$  \\\hline    
$l_L$ &   $(1, 2)$  &  $I_{b e} = -1$ &  
$0$ & $-1$ & $0$ & $0$ & $1$ & $-1/2$  \\\hline    
$N_R$ &   $2(1, 1)$  &  $I_{cd} = 2$ &  
$0$ & $0$ & $1$ & $-1$ & $0$ & $0$  \\\hline    
$E_R$ &   $2(1, 1)$  &  $I_{c d^{\ast}} = -2$ &  
$0$ & $0$ & $-1$ & $-1$ & $0$ & $1$   \\\hline
  $\nu_R$ &   $(1, 1)$  &  $I_{c e} = 1$ &  
$0$ & $0$ & $1$ & $0$ & $-1$ & $0$ \\\hline
$e_R$ &   $(1, 1)$  &  $I_{c e^{\ast}} = -1$ &  
$0$ & $0$ & $-1$ & $0$ & $-1$  & $1$ \\\hline    
\hline
\end{tabular}
\end{center}
\caption{\small Low energy {\em chiral} fermionic spectrum of the five stack 
string scale 
$SU(3)_C \otimes  
SU(2)_L \otimes U(1)_a \otimes U(1)_b \otimes U(1)_c 
\otimes U(1)_d \otimes U(1)_e $, type I D6-brane model together with its
$U(1)$ charges. Note that at low energies only the SM gauge group 
$SU(3) \otimes SU(2)_L \otimes U(1)_Y$ survives.
\label{spectrum8}}
\end{table}

There are various gauged low energy symmetries in the models. They are defined
in terms of the U(1) symmetries Q$_a$, Q$_b$, Q$_c$, Q$_d$, Q$_e$, where the baryon number B and lepton number L, 
respectively are equal to
\beq
Q_a = 3B, \ L = Q_d + Q_e, \  Q_a - 3Q_d -3Q_e = 3(B - L), \ 
Q_c = 2 I_{3R}
\eeq	
and $I_{3R}$ being the third component of weak isospin and
$3 (B-L)$ and $Q_c$ are free of triangle anomalies. The $U(1)_b$ symmetry
plays the role of a Peccei-Quinn symmetry, having mixed
SU(3) anomalies. 

\subsection{The embedding on a string construction} 

The interpretation of the DE in the context of string theory D-brane models, ideally,  would be 
phenomenologically interesting if the string scale is
low at the TeV region.   
At low scale D-brane models (LCD), extra dimensions transverse to the space where the D-branes are wrapping become large, when the string 
scale
becomes low of order $\cal O$(TeV) and at the same time the Planck scale remains large \cite{extra1}, \cite{extra2}. LCD's don't need 
supersymmetry at the TeV and string constructions of low scale D5-brane models appeared in \cite{kokoD5}, \cite{ibaD5}. Also LCD's possess low scale string 
excitations, a 
possible signal for LHC searches \cite{lusttaylor}
and also extra $Z^{\prime}$ gauge bosons as the 
mass of Z' could be at the TeV region, e.g. $M_Z= (g_2^2 + g_Y^2)^{1/2} \upsilon/2 + {\cal O}(\upsilon^2/M_s^2)$ \cite{celis, ir}.
The present toroidal 5-stack D6-brane models don't have large extra dimensions, but one can imagine a scenario that the six-torus can be kept 
small but connected 
to a large volume manifold \footnote{A light Z' gauge boson could be also achieved if 
extra branes are added to the RR tadpoles that do not intersect with the SM branes \cite{feng}.}. 
Recently, assuming a low scale scenario, it has been shown that for a range of string scales between [10-20] TeV, the 5-stack D6-models predict the lowest 
Z' 
excitation
to be in the range [3.5-5.5] TeV while accommodating current anomalies in $b \rightarrow s l^{+} l^{-}$ anomalies \cite{celis}.
\newline
Lets us embed the SM quiver stucture of table (\ref{spectrum8}) in a string compactification.
Our SM quiver can be embedded in a bottom-up approach in string compactification of IIA theory on a six dimensional 
torus equipped with an orientifolded symmetry which converts, 
strings with D9-branes with fluxes compactified on a six-dimensional orientifolded torus T6
(where
internal background gauge fluxes on the branes are turned on) by the  use of a T-duality
transformation on the x4, x5, x6, directions, into
D6-branes intersecting at angles.  In detail,  swe assume that the D$6_a$-branes are wrapping 1-cycles 
$(n^i_a, m^i_a)$ along each of the ith-$T^2$
torus of the factorized $T^6$ torus, namely 
$T^6 = T^2 \times T^2 \times T^2$.
That means that we allow our torus to wrap factorized 3-cycles, that can 
unwrap into products \footnote{We define the homology of the 3-cycle as 
$\Pi_a= \prod_{i=1}^3 (n_a^i [a_i] + m_a^i [b_i])$.
Because of the orientifold $\Omega$R symmetry, where $\Omega$ is the worldvolume parity and R is the reflection
on the T-dualized coordinates,
$T(\Omega R)T^{-1} = \Omega R$, 
each D6$_a$-brane 3-cycle, must have its $\Omega$R orientifold image 
 $\Pi_{a^{\star}}= \prod_{i=1}^3 (n_a^i [a_i] - m_a^i [b_i])$. } 
of three 1-cycles $\Pi_a$, one for each $T^2$.
We define the number of chiral fermions that are located at intersections of the branes a, b to be equal to the homology product of 
3-cycles as
\beq
I_{ab} =\   [\Pi_a] \cdot [\Pi_b] = \prod_{i=1}^3 
( n_a^i m_b^i - m_a^i n_b^i)
\label{homo1}
\eeq
and transforming in the bifundamental representation $(N_a, {\bar N}_b)$ for a left handed fermion with ${I_ab}>0$.   
We also define the intersection number that determines the number of chiral fermions at the intersection of the a-brane with the 
orientifold image of the 
b-brane, namely  
\beq
I_{a b^{\star} } =\  [\Pi_a] \cdot [\Pi_{b^{\star}}] = -\prod_{i=1}^3(n^i_a [a_i] + m^i_a[b_i]) 
\label{homo2}
\eeq
The fermions at the ab* intersection transform at the representation $(N_a, {N}_b)$ for a left handed fermion with ${I_ab*}>0$. 
Any vacuum derived from the previous intersection constraints  
is subject additionally to constraints coming from RR tadpole cancellation 
conditions \cite{ralph1}. That demands cancellation of 
D6-branes charges, wrapping on three cycles with
homology $[\Pi_a]$ and O6-plane 7-form
charges wrapping on 3-cycles  with homology $[\Pi_{O_6}]$. 
Note that the RR tadpole cancellation conditions which represent the
cancellation of RR charges in homology are
\beq
\sum_a N_a [\Pi_a]+\sum_{\alpha^{\prime}} 
N_{\alpha^{\prime}}[\Pi_{\alpha^{\prime}}] - 32
[\Pi_{O_6}]=0.
\label{homology}
\eeq  
These conditions in string theory are stronger that the cancellation of non-abelian gauge anomalies of gauge theories, as it takes into account all the ultarviolet completion of the spectrum of the theory. 
The intersection numbers of the Standard Model quiver of table (\ref{spectrum8}) and figure (\ref{branes}) are solved by the wrapping numbers 
seen at table (\ref{spectrum10}).   
\begin{table}
\begin{center}
\begin{tabular}{||c||c|c|c||}
\hline
\hline
$N_i$ & $(n_i^1, m_i^1)$ & $(n_i^2, m_i^2)$ & $(n_i^3, m_i^3)$\\
\hline\hline
 $N_a=3$ & $(1/\beta^1, 0)$  &
$(n_a^2,  \epsilon \b^2)$ & $(3, {\tilde \epsilon}/2)$  \\
\hline
$N_b=2$  & $(n_b^1, -\epsilon \b^1)$ & $(1/\beta^2, 0)$ &
$({\tilde \epsilon}, 1/2)$ \\
\hline
$N_c=1$ & $(n_c^1, \epsilon \b^1)$ &
$(1/\beta^2, 0)$  & $(0, 1)$ \\    
\hline
$N_d=1$ & $(1/\beta^1, 0)$ &  $(n_d^2,  2 \epsilon \b^2)$  
  & $(1, -{\tilde \epsilon}/2)$  \\\hline
$N_e = 1$ & $(1/\beta^1, 0)$ &  $(n_e^2,   \epsilon \b^2)$  
  & $(1, -{\tilde \epsilon}/2)$  \\
\hline
\end{tabular}
\end{center}
\caption{\small
D6-branes wrapping numbers giving rise to the 
standard model gauge group and chiral spectrum of table (\ref{spectrum8}) at low energies.
The solutions depend 
on five integer parameters, 
$n_a^2$, $n_d^2$, $n_e^2$, $n_b^1$, $n_c^1$,
the NS-background $\beta^i$ and
the phase parameters $\epsilon = \pm 1$, ${\tilde \epsilon} = \pm 1$.
\label{spectrum10}}          
\end{table}

Mixed U(1) gauge anomalies are cancelled via a generalized Green-Schwarz mechanism involving closed sector RR fields. 
Most of the five U(1)'s but 
two of them become massive(see eqn's \ref{rr1}-\ref{rr2}). The massless U(1)'s are  
\beq
  Q^M = n_c^1 (Q_a -3 Q_d -3 Q_e)
-\frac{3 {\tilde \epsilon}\b^2 ( n_a^2 + n_d^2 + n_e^2)}{2 \b^1} Q_c  , \ (3 n_a^2 + 3n_d^2 + 3 n_e^2) \neq 0 ,
\label{hyper}
\eeq
and 
\beq
Q^N =\ \frac{3{\tilde \epsilon}\b^2}{2 \b^1}(Q_a -\ 3Q_d -\ 3Q_e) + 19 n_c^1 Q_c .
\label{extra}
\eeq
When the condition 
\beq
n_c^1 = \ \frac{{\tilde \epsilon}\b^2}{2\b^1}( n_a^2 +\  n_d^2 +\  n_e^2), \ \ n_c^1 \neq 0
\label{mashyper}
\eeq
is satisfied,  Q$^M$ coincides with the 
Standard Model hypercharge assignment 
\beq
U(1)^Y =
\frac{1}{6}U(1)_a -\frac{1}{2}U(1)_c -\frac{1}{2}U(1)_d -\frac{1}{2}U(1)_e \ . 
\eeq 
The tadpole conditions are 
\beq
\frac{9 n_a^2}{ \b^1} +\ 2 \frac{n_b^1}{ \b^2} +\
\frac{n_d^2}{ \b^1} +\ \frac{n_e^2}{ \b^1} +\
N_D \frac{2}{\b^1 \b^2} =\ 16.
\label{ena11}
\eeq
which simply adds a further constraint on the undetermined parameters of table (\ref{spectrum10}). We have allowed the addition of N$_D$ extra branes with 
wrappings $(1/\beta_1, 0)(1/\beta_2, 0)(2, m_D)$ that don't intersect with the rest of the branes and thus don't 
generate additional SM particles.  
  The electroweak Higgses that are necessary for giving masses to all quarks and leptons  are listed in table (\ref{hig}).
\begin{table} [htb] \footnotesize
\renewcommand{\arraystretch}{1}
\begin{center}
\begin{tabular}{||c|c||c|c|c||}
\hline
\hline
Intersection & EW breaking Higgs & $Q_b$ & $Q_c$ &Y \\
\hline\hline
$b c$ & $h_1$  & $1$ & $-1$&1/2\\
\hline
$b c$  & $h_2$  & $-1$ & $1$&-1/2 \\\hline\hline
$b c^{\star}$ & $H_1$  & $-1$ & $-1$& 1/2\\ \hline
$b c^{\star}$  & $H_2$  & $1$ & $1$ & -1/2\\
\hline
\hline
\end{tabular}
\end{center}
\caption{\small 
Electroweak symmetry breaking Higgses in the 5-stack D6-brane model \cite{kokoD6}.
\label{hig}}
\end{table}
Proton is stable due to the fact that baryon number B( is an unbroken gauged
global symmetry surviving at low energies) anomalies cancel through a
generalized Green-Schwarz mechanism(see ).
In the D6-brane models with four stacks \cite{imr} and in the five stack models\cite{kokoD6}, it was noted that the Higsses arising
from intersections bc, bc*, are part of the massive spectrum. They could become massless by varying the 
distance between the parallel branes across the 2nd tori for both intersections. In this work, we follow a different approach. 
The Higgs, from the NS 
sector,  
responsible for electroweak symmetry breaking will be generated by demanding that N=1 supersymmetry is preserved at the intersections 
bc, bc* not necessarily being
the same one. Since the models are non-supersymmetric (non-SUSY) we expect that some intersections to respect some supersymmetry 
providing us with further constraints on the parameters $n_b^1$, $n_c^1$, etc. Related issues on N=1 supersymmetries on non-SUSY D-brane models have been discussed 
in \cite{iba1}, \cite{iba2}.  
The Yukawa interactions for the chiral spectrum of the SM's yield:
\beqa
Y_j^U Q_L U_R^j h_1 + Y_j^D Q_L D_R^j H_2 \ & + &
Y_{ij}^u q_L^i U_R^j H_1 + Y_{ij}^d q_L^i D_R^j h_2 \ + \nonumber\\
Y_h^l \ l_L^{\bar{h}} \ {\nu}_R^{\bar{h}} \ h_1 + Y_h^e l_L^{\bar{h}} e_R^{\bar{h}} H_2 \  &+&
Y_{ij}^N L^i N_R^j h_1 + Y_{ij}^E L_i E_R^j H_2 +\ h.c   \nonumber\\
\label{era1} 
\eeqa
where $i=1, 2$, $j=1,2,3$, $\bar{h}=1$.
The nature of Yukawa couplings is such that the lepton and neutrino sector of 
the models distinguish between different generations, e.g. the first generation from 
the other two generations, as one generation
of neutrinos (resp. leptons)
is placed on a different intersection from the other two one's.
For example looking at the charged leptons of table (\ref{spectrum8}) we see that
one generation of charged leptons $l_L$ gets localized on $be$-intersection,
while the other two generations of leptons $L$ get localized in the $bd$-intersection.


\section{Interpretation of the Diphoton Excess }
\label{sec3}

\subsection{Di-photon preliminaries}


We will consider only the case that the 750 GeV spin-0  resonance  is produced from gluon decays into photons. In this case,
the diphoton excess is explained via the resonant process  pp $\rightarrow$ S $\rightarrow \gamma \gamma$ where S is a new
uncoloured scalar boson with mass M, spin J, and width $\Gamma$ coupled to partons inside the proton. The signal cross section for the scalar mediated process
 mediating on shell scalar singlet $S$, is approximated as follows:
\begin{align}
\sigma(pp \rightarrow S \rightarrow \gamma \gamma) =~& \frac{2J+1}{s M_ \Gamma} ( C_{gg} \Gamma(S \rightarrow gg) + \sum_q C_{q\bar{q}} \Gamma (S \rightarrow q \bar{q}) ) \Gamma(S \rightarrow \gamma \gamma)\,.
\end{align}
Assuming a spin-zero particle produced resonantly via gluon fusion, we arrive at (the last part of the following eqn. can be seen at 
\cite{bena})
\begin{eqnarray}
\sigma(pp \rightarrow S \rightarrow \gamma \gamma) \stackrel{13 \ TeV}{\approx}K_{13} \cdot 4.92 \cdot 10^{6}  \ fb \ 
\frac{\Gamma_{gg}}{\Gamma} \frac{\Gamma_{\gamma \gamma}}{\Gamma} \frac{\Gamma}{M_x} \approx K_{13} \times 
4.9  \cdot 10^{6}   fb  &
\frac{\Gamma_{gg}}{\Gamma} \frac{\Gamma_{\gamma \gamma}}{\Gamma} \frac{\Gamma}{M_x}\nonumber\\
\label{qcd}
\end{eqnarray}
where we have taken into account the QCD NLO enhancement K-factors $K_{13} \approx 1.5$ \cite{fra1, strumia1, strumia2} and
 $\Gamma_{gg} = \Gamma(S\rightarrow gg)$, $\Gamma_{\gamma\gamma} = \Gamma(S\rightarrow \gamma\gamma)$. 
 $\Gamma = \Gamma_{gg}+ \Gamma_{\gamma\gamma}, \sqrt s$ are the total width and the center of mass energy ($\sqrt s=13$ TeV) 
respectively and $C_{gg}$ is 
 the partonic integral~\cite{Martin}
\[  C_{gg}=\int_{M_S/s}^1 f_g(x)f_g( \frac{{\tiny M_S}}{sx})\frac{dx}{x}\,,   \]
\label{parto}
where $f_g(x)$ is the function representing the gluon distribution  inside the proton. 
The integral is computed using MSTW2008NNLO~\cite{Martin}
and its numerical value at $13$ TeV is estimated~\cite{fra1} to be $C_{gg}=2137$.
Thus in the narrow width approximation
 \[ \sigma(gg\to S\to\gamma\gamma)
  =K_{13} \times \frac{1}{M_S\cdot\Gamma\cdot s}C_{gg}\Gamma(S \to gg)\Gamma(S\to\gamma\gamma)\,,\]
The partial widths $ \Gamma(S\to gg),\Gamma(S\to\gamma\gamma)$ 
from loops involving fermions and scalars  are given by~\cite{oua1}(also\cite{fra1})
\beqa 
\frac{\Gamma(X\to gg)}{M_S}&=&\frac{\alpha_3^2}{2\pi^3} \left|\sum_f C_{r_f}\sqrt{\tau_f}y_f {\bar S}(\tau_f)
           +\sum_s C_{r_s}\frac{A_s}{2M_S} P(\tau_s)\right|^2\,,\label{gluon}\eeqa
\beqa
 \frac{\Gamma(S\to \gamma\gamma)}{M_S}&=&\frac{\alpha^2}{16\pi^3} \left|\sum_fd_{r_f}q_f^2\sqrt{\tau_f} y_f {\bar S}(\tau_f)
            +\sum_s d_{r_s}q_s^2\frac{A_s}{2M_S} P(\tau_s)\right|^2\,,  
\label{gfu}        
\eeqa 
where $a_3=0.1 $, $\alpha=1/128$, $C_r$ is the Dynkin index of the colour representation $(C_r=3$ for the triplet), $d_r$ is its dimension, 
$q_s$ the charge and $\sqrt{\tau_{a}}=\frac{2 m_{a}}{M_S}$, with $a=f,s$ for the fermion and scalar  masses respectively.
The functions $ {\bar S}(\tau), P(\tau)$ are \cite{Shif}
\[ {\bar S}(\tau) = 1+(1-\tau) f(\tau),\; P(\tau)=\tau f(\tau)-1\,,\]
\beqa 
f(\tau)&=&\left\{\begin{array}{ll}
                  \arctan^2\frac{1}{\sqrt{\tau-1}}&\tau>1   \\
                  -\frac 14\left(\log\frac{1+\sqrt{1-\tau}}{1-\sqrt{1-\tau}}-i\pi\right)^2;&\tau\le 1
                  \end{array}
\right.\,.
\eeqa 
In order to show that the D-brane model has the ingredients to explain diphoton excess, we
choose the gluon fusion production cross section (\ref{qcd}) at $\sigma(pp \rightarrow \gamma\gamma) \approx 3 $ fb, which is 
the experimentally favoured value as extracted
from a fit to the preferred cross \footnote{See relevant comment on p.29 of \cite{strumia1}.} sections of Morion data conference 
\cite{skata1}, \cite{skata2}, \cite{skata3}  and consider two
sample cases, one of a broad resonance as favoured by ATLAS ($\Gamma = 45$ GeV) and another one of a narrow resonance 
as favoured by CMS. In this case, repeating the procedure of \cite{fra1} and assuming that production from $\gamma \gamma$ partons can be 
neglected with respect to production from gg, we get
\beq
ATLAS, \  \ BR(S \rightarrow \gamma \gamma) \ BR(S \rightarrow gg) \approx 6.1 \times 10^{-7}
\frac{ M_S}{\Gamma} \stackrel{\frac{\Gamma}{M_S} =0.06, \ \Gamma=45 \ GeV}{\approx} 1.02 \times 10^{-5}
\label{K1}
\eeq
\beq
CMS, \  \ BR(S \rightarrow \gamma \gamma) \ BR(S \rightarrow gg) \approx 6.1 \times 10^{-7}
\frac{ M_S}{\Gamma} \stackrel{\frac{\Gamma}{M_S} =0.00013, \ \Gamma=0.1 \ GeV}{\approx} 4.69 \times 10^{-3}
\label{K3}
\eeq
or equivalently 
\beq
ATLAS, \  \ \frac{\Gamma_{gg}}{M_S}\ \frac{ \Gamma_{\gamma \gamma} }{M_S}\approx
6.1 \cdot 10^{-7} \frac{\Gamma}{M_S} \stackrel{\frac{\Gamma}{M_S} =0.06, \ \Gamma=45 \ GeV} \approx 3.66 \times 10^{-8}
\label{K2}
\eeq
\beq
CMS, \  \ \frac{\Gamma_{gg}}{M_S}\ \frac{ \Gamma_{\gamma \gamma} }{M_S}\approx
6.1 \cdot 10^{-7} \frac{\Gamma}{M_S} \stackrel{\frac{\Gamma}{M_S} =0.00013, \ \Gamma=0.1 \ GeV} \approx 8.13 \times 10^{-11}
\label{K4}
\eeq
where we have neglected the $K_{13}$ factor in the calculation of  (\ref{K1}), (\ref{K2}), (\ref{K3}), (\ref{K4}).


\subsection{The sneutrino ${\tilde \nu}_R$ a candidate for generating the diphoton excess}  


In \cite{kokoD6}  we imposed N=1 supersymmetry (SUSY) on the intersection ce of the intersecting at angles branes c and e, as a means to 
generate the spartner of 
$\nu_R$, namely ${\tilde \nu}_R$, which was used to break the   
 extra U(1) (\ref{extra})
that is surviving massless beyond hypercharge at low energies\footnote{There is no B $\wedge F_i$ Chern-Simons couplings to the hypercharge $Q^M$, where $F_i$ the non-abelian 
field strength of the U(N)$_i$ gauge bosons. As a result $Q^M \equiv Q^Y$ survives massless to low energies. }. The N=1 SUSY 
condition \footnote{We define the complex structure as $U^i = \frac{m_i \ R_i^2 }{n_a \  R_1^i}$, i=1, 2, 3 for the three tori. See 
\cite{kokoD6} for details. } on intersection ce is 
\beq
ce: \ \ +(tan^{-1}\frac{\epsilon \beta_1 U_1}{n_c^1}) +(- tan^{-1}\frac{\epsilon \beta_2 U_2}{n_e^2})-
 (tan^{-1}\frac{{\tilde \epsilon} U_3}{2} + \frac{\pi}{2}) = 0 \ ,
\label{ce}
\eeq
which is solved (possessing the N=1 susy (++-) in the notation of \cite{iba1}) by choosing  
\beq
 \frac{\epsilon \beta_1 U_1}{n_c^1} = \frac{{\tilde \epsilon} U_3}{2}, \ \ -tan^{-1}\frac{\epsilon \beta_2 U_2}{n_e^2} = \frac{\pi}{2} 
\label{rf1}
\eeq
\beq
 \ n_e^2 = 0, \ \epsilon < 0
\label{com1}
\eeq 
That necessarily fixes 
\beq
\epsilon = -1 , \  \tilde \epsilon = 1, \ n_c^1 < 0
\label{rf2}
\eeq
In this work,  we interpret the observed diphoton excess as a 750 GeV scalar neutrino ${\tilde \nu}_R$ resonance generated via gluon fusion. Since it has a low mass,
it can no longer be used to break the U(1) (\ref{extra}). Instead this U(1) could be broken by another gauge singlet scalar, the 
spartner of $N_R$'s located at the intersection cd which becomes massless if there is a N=1 SUSY preserved at the intersection cd
(namely the ++-, N=1 susy in the notation of \cite{iba1}). The latter happens when the condition
\beq
cd : \ \ +(tan^{-1}\frac{\epsilon \beta_1 U_1}{n_c^1})+(tan^{-1}\frac{2 \epsilon \beta_2 U_2}{n_d^2})-(\frac{\pi}{2}+tan^{-1}(\frac{{\tilde \epsilon}U_3}{2})) = 0 \ 
 \stackrel{(3.3)}{\rightarrow} \ n_d^2 = 0
\label{com2}
\eeq
is satisfied, where we have also used (\ref{rf1}). Since $\nu_R$ is located with multiplicity one at an intersection, while N$_R$ appears with multiplicity two at 
another intersection, one of the spartners of $N_R$'s could be used to break the U(1) (\ref{extra}). Also 
we could (as was applied recently at \cite{celis}) identify $\nu_R$ as belonging to the third generation where also we identify Q to be (t, b)$_L$ and  identify q to be (u, d)$_L$, (c, s)$_L$; we also identify L to represent  (e, $\nu_e$)$_L$ and 
($\mu$, $\nu_µ$)$_L$ and $l_L$ to be ($\tau$, $\nu_{\tau}$ )$_L$; also U, D represent (u, c, t)$_R$ and (d, s, b)$_R$
respectively. Also (e, $\nu$)$_R$ is identified with the right-handed electron and electron-neutrino while (E, N)$_R$
are identified with the first and and second generation right-handed leptons.
Thus our 5-stack D6-brane model at the string scale generates beyond the chiral spectrum of the SM and also the spartners of the neutrinos, 
by imposing some N=1 SUSY to be preserved at the ce intersection. There are also non-chiral fermions at the intersections ad, ad*, 
ae, ae*, bc, bc* de, be* as it happens some branes to be parallel along the same $T^2$ torus and intersecting at angles along the rest of the 
tori. We discuss later those issues as matter it concerns their effect on DE.


\subsection{Using fermions to saturate the decay widths {\small $ S \rightarrow \gamma \gamma  , S \rightarrow  g g $} }
\label{gluon1}

$\bullet$ {\em Contributions to $\Gamma( S \rightarrow  g g)/M_x $ from coloured vector-like fermion states}
\newline
\newline
The ${\tilde \nu}_R$ (identified as S from now on) generates the Yukawa coupling interaction\footnote{The intersection numbers inside the parenthesis denotes
the corresponding fermion that is localized at this intersection.}
\beqa
y \ S_{(0, 0, 1, 0, -1)} \ (I^{ac*}_{(-1, 0,-1, 0, 0)} )\ (I^{ae*}_{(-1, 0, -1, 0, 0)} )   = y\  S^0 \ D_R^{1/3} \ {X_1}^{-1/3} 
\label{ea1}
\eeqa
is the only\footnote{The superscript denotes the hypercharge assignment.} fermionic contribution to the gluon fusion partial width $ \Gamma(S \to gg)$. $X_1$ 
is located at the intersection ae*. Because $I_{ae*} = 0$ at this intersection, we have generation of the non-chiral coloured states $(X_1)_{(1,0,0,0,1)}=(3,1), (X_2)_{(-1,0,0,0,-1)} =
 ({\bar 3}, 1)$. The number of $X_1$'s is calculated by the non-zero intersection number in the non-parallel tori and is given by
 $I_{X_1}=|\beta_2(n_e^2 + n_a^2) |\stackrel{\ref{com1}}{=}|\beta_2 n_a^2| $. Choosing (for reasons clarified, later on, in the paper)
\beq
\beta_2=1, \  \ n_a^2 = -2
\label{con1}
\eeq
and substituting $n_e^2$ from (\ref{com1}), we have at least 2 $X_1$'s generating 2 vector pairs by pairing them with the light quarks $ D_R = b_R$. Generally this mixing could influence proton decay. However, in our models baryon number is a gauge symmetry and thus proton is stable.  
The contribution of $b_R$ with a mass of $m_{b_R} = 4.7$ MeV to (\ref{gfu}) is small, of the order of $10^{-10} y^2$ (assuming  $y\approx {\cal O} (1)$) and 
may not be considered further\footnote{The other two generations of right handed down quarks, d, s provide us with small contributions as well that can 
be neglected.}  
\begin{table}
\begin{center}
\begin{tabular}{||c||c||c||}
\hline
$\sharp$ X$_1$  & X$_1$ \ Mass & $\Gamma_{gg}$/M$_x$ \\
\hline
 $2(3,1)$ & $800$  &
$1.7 \times 10^{-5} y^2$ \\
\hline
$2(3,1)$  & $900$ & $1.3 \times 10^{-5} y^2$ \\
\hline
$2(3,1)$ & $1000$ & $ 1.0 \times 10^{-5} y^2$\\
\hline
$2(3, 1)$  & $1100$ &$8.8 \times 10^{-6} y^2$ \\    
\hline
$2(3,1)$ & $1200$ &  $7.3 \times 10^{-6} y^2$   \\
\hline
$2(3,1)$ & $1500$ &  $4.6 \times 10^{-6} y^2$   \\
\hline
$2(3,1)$ & $2000$ &  $2.6 \times 10^{-6} y^2$   \\
\hline
$2(3,1)$ & $2200$ &  $2.1 \times 10^{-6} y^2$   \\
\hline
$2(3,1)$ & $2500$ &  $1.6 \times 10^{-6} y^2$   \\
\hline
$2(3,1)$ & $3000$ &  $1.1 \times 10^{-6} y^2$   \\
\hline
$2(3,1)$ & $3100$ &  $1.0 \times 10^{-6} y^2$   \\
\hline
$2(3,1)$ & $3200$ &  $9.9 \times 10^{-7} y^2$   \\
\hline
\end{tabular}
\end{center}
\caption{\small
Contributing exotic bottom quark with charge 1/3 coloured state from eqn.(\ref{ea1}) to the $\Gamma_{gg}/M_x$.
\label{spee1}}          
\end{table}
Limits on the masses of vector-like bottom B quarks with electric charge -1/3  and vector couplings to W, Z, and H bosons do exist at 95 $\%$ confidence level, assuming decays into standard model particles such as $B \rightarrow W t, Zb, Hb$ at
95 $\%$ C.L. Lower limits are in the range 740-900 GeV \cite{bot1} or 575-813 GeV \cite{bot2} for CMS and
ATLAS respectively. In table (\ref{spee1}) we are summarizing the contributions of $X_1$ exotic bottom quarks with masses between
 800 GeV and 3200 GeV
to $\Gamma_{gg}/M_S$. We consider their Yukawa coupling a free parameter and use $C_{X_1} = 1/2$ in (\ref{gluon}). skata
We note that perturbative Type IIA string theory allows in general
for a range of $y$ including large values in the interval $y \in [O(1), O(50)]$ \cite{cve1}. 
\newline\newline
$\bullet$ {\em Contributions to $\Gamma( S \rightarrow  \gamma \gamma)/M_S$ } 
\newline\newline
Contributions to the $ S \rightarrow  \gamma \gamma $ loop diagrams 
could come from the Yukawa mass terms involving mixing of S with the right handed fermion singlets from cd* intersection, the $E_R$ and its vector pair from 
de* intersection as follows from skata
\beq
y_s \cdot S_{(0,0,1,0,-1))}^{Y=0} \cdot (I_{cd*})^{Y=+1}_{(0,0,-1,-1,0)} \cdot (I_{de*})^{Y=-1}_{(0,0,0,1,1)} = y_s  \cdot S \cdot E_R^{Y=1} \cdot (I_{de*})^{Y=-1}_{(0,0,0,1,1)} 
\label{no}
\eeq
However, the number of  $I_{de*}$ fermions is equal to  
$\beta_2 (2 n_d^2+n_e^2) \stackrel{ (\ref{com1}, \ref{com2})  }{=}0$.  
Thus there is no contribution from (\ref{no}) to $ S \rightarrow  \gamma \gamma $ width.
The following vector pair of fermion weak doublets seen in the Yukawa coupling 
\beq
y_ {\tilde h} \ S \cdot (I_{bc})^{1/2} \cdot  (I_{be})^{-1/2} = y_{\tilde h} \ S \cdot (I_{bc})^{1/2} \cdot l_L^{-1/2} = 
\label{dou1} y_{\tilde h} \ S^0 \cdot h_1^{1/2} \cdot l_L^{-1/2} \ , 
\eeq
contributes the to $ S \rightarrow  \gamma \gamma $ width. It is generated by a mixing from the higgsinos ${\tilde h}_1$ from bc 
intersection with the tau lepton, the lepton doublet of the third generation, that appears with multiplicity one.
The number of higgsinos \footnote{At the bc intercection, we have localized a non-chiral pair of 
Higgsinos, each of them appearing with multiplicity given by (\ref{as1}) and (\ref{asa1}). Explicitlty, they are 
${({\tilde h}_1)}_{(0, \ 1, \ -1, \ 0, \ 0)}$, ${({\tilde h}_2)}_{(0, \ -1, \ 1, \ 0, \ 0)}$.
} 
is calculated from the non-zero intersection numbers in the first and third tori
\beq
|I_{bc}| = |(\epsilon\beta_1)(n_b^1+n_c^1 )|, \ n_c^1 = \frac{\beta_2}{2 \beta_1}n_a^2
\label{as1}
\eeq  
Making the choice \footnote{The choice $n_b^1 = 0$ may be justified later. See (\ref{sca1}).}
\beq
n_a^2 \stackrel{(\ref{con1})}{=} -2, n_c^1 \stackrel{\ref{as1}}{=} -2, \  n_b^1 = 0, \ \beta_1 =1/2, \ \beta_2 \stackrel{(\ref{con1})}{=}1 \rightarrow
 I_{bc} = 1
\label{asa1}
\eeq
the number of higgsinos is one, thus generating one vector weak pair of doublets that mixes with the tau lepton.
Using the tau lepton mass $m_{\tau} = 1.78 $ GeV, we find that its contribution to the decay width 
$\Gamma(S \rightarrow  \gamma \gamma)/M_S $ is small of the order of $6 \times 10^{-12}$ and can be neglected in 
the following. 
In the case in question, colour triplets with charge 1/3 are mediating $S\rightarrow \gamma \gamma$ with mass $M_c$, Yukawa coupling 
$y$, $Q_c = 1/3$,  
$d=2$,  while  ${\tilde h}_1$ higgsinos with mass $M_{\tilde h}$ and coupling  $y_{\tilde h} $ contribute with $Q_{\tilde h} = 1$, d=1. 
At table \ref{spe2}, we present the results of fixing the value of higgsinos at 800 GeV and subsequently calculating their contribution to $\Gamma_{gg}/M_S$ 
assuming the 
Yukawa couplings of the triplets are in the perturbative regime $\cal{O}$(1) \cite{cve1}
\beq
y_s = 1, m_{\tilde h} = 800  \ GeV       
\eeq
Then by varying the values of the colour triplet between 800-3200 GeV, we calculated 
the values of the higgsino couplings, such that the product $\Gamma_{gg} \Gamma_{\gamma \gamma}/M_S^2$ of eqn. (\ref{K4}) 
takes its 
LHC predicted value at 
$\sigma (pp \rightarrow S \rightarrow \gamma \gamma ) \approx$ 3 fb assuming a narrow width $\Gamma_{total} =0.1$ GeV. 
Table (\ref{spe2}) summarizes our results.       
\begin{table}
\begin{center}
\begin{tabular}{|c|c|c|c|c|c|}
\hline
$y_s$ &  $y_{\tilde h}$ & $M_c \equiv M_{X_1}$  & $\Gamma_{g g}/M_S$ & $\Gamma_{\gamma\gamma}/M_S$ & $m_{\tilde h}$\\
\hline
$1$ &  $17.88$ & $800$  & $1.7 \times 10^{-5} y^2$ & $4.78  \times 10^{-6}$    & $800$ \\
\hline
$1$ & $20.71$  & $900$ & $1.3 \times 10^{-5} y^2$ &  $ 6.25  \times 10^{-6} $     & $800$\\
\hline
$1$ & $23.84$ & $1000$ & $ 1.0 \times 10^{-5} y^2$&   $  8.13  \times 10^{-6} $   &$800$\\
\hline
$1$ & $25.49$  & $1100$ &$8.8 \times 10^{-6} y^2$ &     $0.92  \times 10^{-5} $    & $800$\\    
\hline
$1$ & $28.20$ & $1200$ &  $7.3 \times 10^{-6} y^2$   &    $1.11  \times 10^{-5} $    &$800$\\
\hline
$1$ & $35.80$ & $1500$ &  $4.6 \times 10^{-6} y^2$   &    $1.77  \times 10^{-5}  $    &$800$\\
\hline
$1$ & $47.90$ & $2000$ &  $2.6 \times 10^{-6} y^2$   &    $3.13  \times 10^{-5} $    &$800$\\
\hline\hline
$1$ & $53.40$ & $2200$ &  $2.1 \times 10^{-6} y^2$   &  $ 3.87  \times 10^{-5} $    &$800$\\
\hline
$1$ & $61.25$ & $2500$ &  $1.6 \times 10^{-6} y^2$  &    $5.08  \times 10^{-5} $    &$800$\\
\hline
$1$ & $74.00$ & $3000$ &  $1.1 \times 10^{-6} y^2$   &    $7.39  \times 10^{-5} $    & $800$\\
\hline
$1$ & $77.60$ & $3100$ &  $1.0 \times 10^{-6} y^2$  &    $8.13  \times 10^{-5}$   &$800$\\
\hline
$1$ & $78.00$ & $3200$ &  $9.9 \times 10^{-7} y^2$   &     $0.84  \times 10^{-4} $ &$800$\\
\hline
\end{tabular}
\end{center}
\caption{\small
Values of Yukawa couplings for the higgsino and the exotic charge 1/3 quark are obtained within the perturbative regime of 
string Yukawa couplings in the intersecting D6-brane model \cite{kokoD6}. We assume $\Gamma_{total} = 0.1$ GeV.
\label{spe2}}          
\end{table}
We find that, at $\sigma \approx 3$ fb, with fixed higgsino mass $m_{\tilde h} = 800$ GeV,  only values of the exotic down quark with charge 1/3, between [800-2000] GeV, possess 
perturbative string couplings (excluding the last five entries of table (\ref{spe2})) that are consistent with the CMS predicted 
diphoton excess rates (\ref{K4}).

\begin{figure} 
\begin{center}
\includegraphics[width=.8\textwidth]{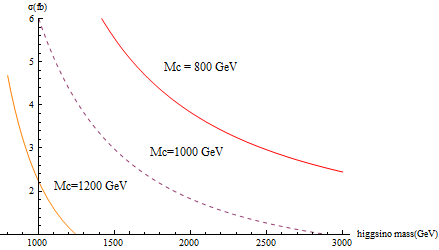}
\end{center}
\caption{The cross section $\sigma (pp \rightarrow X \rightarrow \gamma \gamma)$ (in fb units) in the parametric space of the
Higgsino $h_1$ for a selection of the exotic down quark masses with charge 1/3  at 800, 1000, 1200 GeV. The 
couplings take values within perturbative string theory regime. We have set  $y_s =1$ and $y_{\tilde h} = 12$. }
\end{figure}

\subsection{Diphoton excess in the presence of extra scalars}
\label{gluon2}

We have shown that a scalar superpartner of the right handed neutrino can account for the diphoton excess signal of 750 GeV 
and also discussed the 
presence of vector-like fermions
which populate eqn.'s (\ref{gluon}, \ref{gfu}) to explain the diphoton excess. 
It is also possible to generate vector-like scalar contributions to eqn.'s (\ref{gfu}) by demanding that N=1 supersymmetry is 
preserved at the intersections $I_{ij}$ where the vector-like fermions are localized. Thus the previously massive scalars become massless
at the intersections completing the N=1 chiral multiplet structure. The scalar may appear with the same multiplicity as the corresponding fermion. 
Wc will briefly discuss the conditions for these scalars to exist, without further providing further details of their contribution to 
$\Gamma (pp \rightarrow S \rightarrow \gamma \gamma)$. 
\newline\newline
$\bullet$ {\em Contributions to $\Gamma( S \rightarrow  g g)/M_x $ from coloured vector-like scalar states}
\newline
 Scalar contributions to  $\Gamma( S \rightarrow  g g)/M_S$ could be generated  by demanding that in eqn.(\ref{ea1}), the intersections
bc, be a N=1 supersymmetry is preserved by the corresponding branes and thus the non-chiral spartners of higgsinos the electroweak Higgses and the spartners of 
the the 3rd generation of leptons appear respectively.    
Note that as the models we discuss are not supersymmetric, these supersymmetries need not be necessarily the same.
The N=1 supersymmetry condition at the intersection\footnote{At the bc intersection there are two N=1 supersymmetries that are 
preserved, namely the (+++), (+-+). Thus this is a N=2 supersymmetry preserving intersection. Necessarily, the conditions (\ref{rf1}), (\ref{sca1}) also solve 
the supersymmetry condition on the intersection bc*, that is 
\beq
(-tan^{-1}\frac{ {\epsilon} \beta_1 U_1}{n_b^1}  + tan^{-1}\frac{\epsilon \beta_1 U_1}{n_c^1} ) \pm 0 -( tan^{-1}\frac{{\tilde \epsilon}U_3}{2}
+\frac{\pi}{2}) = 0
\eeq
Note that the pair of  N=1 susy's preserved at bc* intersection, namely the (++-), (+--), are different than the one's preserved at the bc intersection.
} bc is \footnote{Where the expressions inside the parenthesis denote the angles in the respective complex 3-dimensional orientifolded tori.}
\beq
bc \ : \  (-tan^{-1}\frac{\epsilon \beta^1 U^1}{n_b^1} - tan^{-1}\frac{\epsilon \beta^1 U^1}{n_c^1}) \pm 0 +
(tan^{-1}
\frac{{\tilde \epsilon} U_3}{2} - \frac{\pi}{2}) = 0 \ ,
\label{bc}  
\eeq
which is solved by eqn. (\ref{rf1}) and 
\beq
 - \frac{\epsilon \beta_1 U^1}{n_b^1} \ = \ \frac{\pi}{2} \rightarrow \ \ n_b^1 = 0, \  \ \epsilon \stackrel{ (\ref{rf1})}{=} -1
\label{sca1}
\eeq
Condition (\ref{bc}) generates the electroweak scalar Higgses, superpartners of higgsinos ${\tilde h}_1$, ${\tilde h}_2$, seen at table (\ref{hig}).
 
The N=1 supersymmetry condition at the intersection be (generating the scalar superpartner of $l_L$)  is 
\beq
be: \ \ +(-tan^{-1}\frac{\epsilon \beta_1 U^1}{n_b^1}) +( - tan^{-1}\frac{\epsilon \beta_2 U^2}{n_e^2}) - 
(tan^{-1}\frac{{\tilde \epsilon} U_3 }{6} + tan^{-1}\frac{{\tilde \epsilon}U_3}{2})=0 \ , 
\eeq
which is solved  by 
\beq
tan^{-1}\frac{{\tilde \epsilon} U_3 }{6} + tan^{-1}\frac{{\tilde \epsilon}U_3}{2} =  \pi, \ \ {\tilde \epsilon} = 1
\label{sol12}
\eeq
Solving (\ref{sol12}) the value of complex structure moduli across the third torus is fixed at 
\beq
U_3 \approx 4.2 \times 10^{16}
\label{u3}
\eeq

\subsection {Estimating the mass of ${\tilde \nu}_R$ }

The mass of ${\tilde \nu}_R$ of the yet unseen 750 GeV diphoton excess candidate in our D-brane model can be generated 
geometrically by varying 
slightly the complex structure $U_1$ in the first 
torus. In fact, by setting its mass to 750 GeV, we will be able
to set constraints on the string scale of the models.
This procedure is equivalent to turning on a Fayet-Iliopoulos term 
in the effective theory. 
Assuming a slight departure of $U_1$ from its value  \eqref{com1}
\beq
U^1 = \frac{n_c^1}{2 \beta_1}\frac{{\tilde \epsilon}}{\epsilon} U^3 + \delta_1
\label{del1}
\eeq
where $ \delta_1 << U_3$,
we find ($M_s$ the string scale)
\beq
m^2_{{\tilde \nu}_R} = - \frac{M_s^2}{2} \frac{\frac{2 \beta_1}{n_c^1}\delta_1  }{1+(\frac{ {\tilde \epsilon}U_3}{2})^2 }
\eeq 
As a representative example,  we assume that $m_{\nu_R} = 750$ GeV, the string scale $M_s =2 \times 10^{16}$ GeV and using $\beta_1 = 1/2, \ n_c^1 = -2$, we find that 
\beq
\delta_1 = 2.48 \times 10^6 
\label{comple1}
\eeq
is really a small number compared to $U_3$ value (\ref{u3}). Eqn. (\ref{del1}) also fixes moduli $U_1$.
Varying the string scale from $ 7 \ TeV\  < \ M_s \ < 10^{19}$ GeV, we list at table (\ref{spek1}) that small numbers (less or equal to 1$\%$ of $U_3$) for the variation $\delta_1$ are produced only 
for $ 10^{14}\  < \  M_s \ < \ 10^{19}$ GeV and thus a high string scale is preferred.  
\begin{table}
\begin{center}
\begin{tabular}{||c||c||c|c|c|c|c|c|c|c|c|}
\hline
 $M_s \  (GeV) $   &   $10000$   & $10^{10}$   &  $10^{11}$ &   $10^{12}$   &  $10^{13}$ &  $10^{14}$ & $10^{15}$ & $10^{16}$ & $10^{18}$& $10^{19}$ \\
\hline
$\delta_1 \times 9.92$    &  $10^{30}$    & $10^{18}$   & $10^{16}$  &  $10^{14}$ &  $10^{12}$  &    $10^{10}$  & $10^{8}$& $10^6$& $100$ & $1$\\
\hline
\end{tabular}
\end{center}
\caption{\small
Values of complex structure variation against the string scale.
We assume that the mass of ${\tilde \nu}_R$ equals 750 GeV. The preferred values of $M_s \in \{ 10^{14}-10^{19} \}$ GeV.  Values of $\delta_1$ 
are 
 compared against the value of $U_3$. 
\label{spek1}}          
\end{table}
For the parameter values we have used, namely
$n_e^2 = n_d^2 = n_b^1 = 0$, $ -\epsilon = \  {\tilde \epsilon} = 1, \ \beta_1 = 1/2, \ \beta_2 = 1$
the tadpole condition  (\ref{ena11}) is satisfied for $N_D = 13$ hidden anti-branes. Thus the wrappings of the extra branes become
 $(1/\beta_1, 0)(1/\beta_2, 0)(2, -m_D)$. 
The high string scale result was expected as at the present string models have a naturally high string scale since there are no  
transverse dimensions to the branes that can be made large by lowering the string scale at the TeV \cite{extra1}, \cite{extra2}.

\section{The axion as a string theory 750 GeV candidate responsible for di-photon excess}
\label{sec4}

In \cite{cve2}, \cite{anto1}, \cite{anto3}, \cite{ibaax} the axion was discussed as a solution to the diphoton excess problem in the context of 
string theory.  The parameters of our classes of D6-brane string models, accommodate these scenaria when certain superpartners of SM particles are 
present. 
  
The effective axion Langrangian for an axion ${\Phi}$ coupled to gluons and photons for our toroidal models is (see \cite{ibaax} for relevant dicsussion)
\begin{equation}
\mathcal{L}_{a_0}=\frac{\alpha_s}{4\pi} g_{g}\frac {\Phi}{f} G_{\mu\nu}\widetilde{G}^{\mu\nu} +
 \frac{\alpha_{Y}}{4\pi} g_{Y} \frac {\Phi}{f} B_{\mu\nu}\widetilde{B}^{\mu\nu},
\label{axio}
\end{equation}
where $f$ the axion  decay coupling constant; $\alpha_s$, $\alpha_Y$ the strong and hypercharge fine structure constants; $g_s$, 
$g_{Y}$ are model dependent coefficients.  
In the context of the toroidal models we are discussing there are four potential candidate axions $\Phi^i$, $i=0, 1, 2, 3$. Their duals,   
the RR sector two forms $B_2^i$ (and their four dimensional Poincare dual axion scalars $\Phi^i$) couple to all the U(1) field strengths $F_i$  
\cite{imr}
(part of the original U(N) gauge groups of the different brane stacks) as
\beq
\sum_{\alpha} k^i_{\alpha} \left( B^{i} \wedge tr(F^{\alpha})\right), 
\eeq
In the present models  the five different U(1)'s couple as \cite{kokoD6} : 
\beqa
B_2^1 \wedge \left( \frac{- 2 \epsilon {\tilde \epsilon}    \b^1 }{\b^2 } 
\right)F^b,&\nonumber\\
B_2^2 \wedge \left(\frac{\epsilon \b^2}{\b^1}  
\right)(9F^a + 2   F^d+  F^e),&\nonumber\\
B_2^3  \wedge \left( \frac{3 {\tilde \epsilon} n_a^2}{2\b^1} F^a +     
\frac{n_b^1}{\b^2}F^b  + \frac{n_c^1}{\b^2} F^c -
\frac{{\tilde \epsilon} n_d^2}{2\b^1} F^d
-\frac{{\tilde \epsilon} n_e^2}{2 \b^1}F^e \right).&
\label{rr1}
\eeqa
Notice that $B \wedge F$ couplings induce a Stueckelberg mass term for the anomalous U(1)'s that has a 
non-zero $k^i_a$.   The three U(1)'s that couple to $B_2^i$, i=1,2,3, cancel their triangle anomalies, receive a mass and the 
corresponding U(1)'s remain as global symmetries 
to low energies. The fourth U(1) combination, the hypercharge (\ref{hyper}, \ref{mashyper}, \ref{ena11}) also remains massless as it does not couple to any 
$F^i$'s.  The coupling of $B_2^0$ to any $F^i$ is zero (as we have imposed the condition $\Pi_{i=1}^3 m^i =0$ to all branes by construction 
to this class of 
models) and thus the associated axion $\Phi^0$ that stays massless. 
The associated axions $\Phi^i$ from the closed string sector couple to the U(N) field strengths as  
\beq
 \sum_{\alpha} \lambda^i_{\alpha} \ \Phi^i 
 \ tr ( F^{\alpha}  \wedge F^{\alpha}), \ \alpha = a, b, c, d, e ; \  i =0, 1, 2, 3
\eeq
cancelling the mixed U(1) triangle anomalies $A^{ij}$ of the massive U(1)'s to the gauge groups as  $A^{ij} + k^i_{\alpha} \lambda^j_{\alpha} = 0$. 
In fact, the lagrangian coupling of the axion becomes
\beq
\Phi_o \left(   \lambda_o^a \ F_a \wedge F_a  + \lambda^b_o \  F_b \wedge F_b  +  \lambda_o^d  \  F_d \wedge F_d +
 \lambda_0^e  \  F_e \wedge F_e \right),
\eeq 
where $\lambda_i^a$ are model dependent coefficients. In our models, the non-zero axion-like couplings are 
\beqa
\Phi^1 \wedge [  \frac{\epsilon {\tilde \epsilon}\b^2 }{2 \b^1} (F^a \wedge 
F^a) - \frac{\epsilon {\tilde \epsilon} \b^2 }{ \b^1} (F^d \wedge F^d) -
\frac{ \epsilon  {\tilde \epsilon} \b^2 }{2 \b^1}
 F^e \wedge F^e )], &\nonumber\\
\Phi^2 \wedge [ \frac{- \epsilon \b^1 }{2 \b^2 } (F^b \wedge 
F^b) +  \frac{ \epsilon \b^1 }{ \b^2 }(F^c \wedge F^c)    ],
&\nonumber\\
\Phi^o \wedge \left(  \frac{3n_a^2}{\b^1}(F^a \wedge 
F^a)  +     \frac{{\tilde \epsilon}n_b^1}{ \b^2}(F^b \wedge 
F^b)  + \frac{n_d^2}{ \b^1}(F^d \wedge 
F^d)  + \frac{n_e^2}{ \b^1}(F^e \wedge F^e) \right),&
\label{rr2}
\eeqa
The present models offer a variety of possibilities as matter as it concerns the possible couplings of axion to QCD field strength and the photon. 
We list them as follows :

 \subsection{ Producing the axion via photon fusion} 

In section (\ref{sec3}),  sneutrino was the 750 GeV candidate that was produced via gluon fusion. Assuming at this section that the 
axion (and not the ${\tilde \nu}_R$) could be the 750 GeV candidate, the scalar  ${\tilde \nu}_R$ could be used to 
break the extra beyond hypercharge U(1) (\ref{extra}).
Let	 us further assume that the axion is produced via photon fusion. This possibility has been discussed in 
the four stack models of \cite{anto1} assuming a low scale string theory.
In our 5-stack parametric classes of string models, photon fusion is easily accommodated if the coupling of the axions to the SU(2), SU(3) 
field strengths become zero, namely   
\beqa    
 n_a^2 = 0,  \ n_b^1 = 0, \  n_e^2 = 0 \nonumber\\  (or  \ n_a^2 = 0, \  n_b^1 = 0, n_d^2 = 0)
\eeqa
We have seen in (\ref{sca1}) that the coefficient $n_b^1$, which describes the axion coupling to the SU(2) gauge bosons, can be 
relaxed to zero as a result of imposing N=1 supersymmetry on the intersections $bc, bc*$, 
thus generating the Higgs necessary to give masses to all quarks and fermions.
The condition $n_e^2=0$  is derived in ({\ref{com1}) by demanding that a N=1 supersymmetry is preserved  
at the intersection $ce$, in order to generate a massless sneutrino\footnote{Alternatively, we could have let $n_e^2 \neq 0$, in which 
case, necessarily,
since we have to break the extra U(1) (\ref{extra}),  we will generate the required gauge singlet scalars by enforcing N=1 supersymmetry 
at the de intersection (see 
eqn. (\ref{com2})). This procedure generates the two sneutrinos ${\tilde N}_R$, superpartners of $N_R$, both localized at the same intersection. One of them, could be used to break the 
U(1) (\ref{extra}).} 
that is used to break the U(1) (\ref{extra}).
Finally, the condition $n_a^2=0$ is necessary if we want the axion to be produced by photon fusion. In this case (as in \cite{anto1}, \cite{anto3}),
 the axion should not couple to colour SU(3) gauge bosons 
to avoid unwanted diphoton diphoton signals.  We derive this condition by demanding that the spartner of right handed up quark $U_R$ is generated 
when N=1 supersymmetry is preserved at the intersection ac where the spartner of $U_R$ is localized. The supersymmetry condition at ac is
\beq
ac : \  \ (tan^{-1}\frac{\epsilon \beta_1 U_1}{n_c^1}) + (tan^{-1} \frac{\epsilon \beta_2 U_2}{n_a^2})+(tan^{-1}\frac{{\tilde \epsilon} U_3}{6}-
\frac{\pi}{2})=0
\label{su3}
\eeq
The branes a, c preserve the N=1 supersymmetry (-++). Eqn. (\ref{su3}) is solved by (\ref{rf1}), (\ref{rf2}) and
\beq
tan^{-1}\frac{\epsilon \beta_2 U_2}{n_a^2} = \frac{\pi}{2} - tan^{-1}\frac{{\tilde \epsilon} U_3}{6} - tan^{-1}\frac{{\tilde \epsilon} U_3}{2}
\label{cond1}
\eeq
Eqn. (\ref{cond1}) is solved by 
\beq
n_a^2 = 0
\label{cond2}
\eeq
Using also that $\epsilon$ is negative (see eqn.(\ref{sca1})) results in the condition (\ref{sol12}), that was derived by demanding that 
the spartner of ${l}_L$ is 
generated at the intersection be, fixing the $U_3$ modulus at its value (\ref{u3}).  Let us assume that the same supersymmetry (-++) is preserved 
at the ab* intersection. Then conditions (\ref{cond2}) and also (\ref{sol12}) which solve (\ref{cond2}) at intersection ac, also solve the N=1 
supersymmetry condition at ab*, namely
\beq
ab* : \ \ (tan^{-1}\frac{\epsilon \beta_1 U_1}{n_b^1}) + (tan^{-1}\frac{\epsilon \beta_2 U_2}{n_a^2})+
( tan^{-1}\frac{{\tilde \epsilon} U_3}{6} + tan^{-1}\frac{{\tilde \epsilon} U_3}{2}) = 0
\label{gener1}
\eeq
 Condition (\ref{gener1}) generates the spartner of $q_L$. 
\subsection{Producing the axion via gluon fusion} 

Axionic gluon fusion (AGF) has been discussed in a string theoy context in \cite{cve2}, \cite{ibaax}. The axion 
AGF could be present in our models if a non-zero coupling of axion to the SU(3)$^2$ field strength and the photon $F^2$ exists. Thus we need
\beqa
 n_b^1 = 0, \  \  n_e^2 = 0 \  (or \  n_b^1= 0, \ n_d^2 = 0)
\label{cond}
\eeqa
 In \cite{ibaax} the value of $n_b^1$ was allowed from RR tadpole conditions to be chosen equal to zero. 
In our models the conditions (\ref{cond}) are obtained, as we require the presence of Higgses, higgsinos and 
${\tilde \nu}_R$ simultaneously with the SM quarks and leptons.  As we have already discussed, this is achieved with the simultaneous 
presence of supersymmetry on intersections, bc, bc* and ce respectively.  As seen from  (\ref{axio}), (\ref{rr2}),  the QCD 
coefficient $g_g$ in (\ref{axio}) is proportional to the $n_a^2$ coefficient. Similarly, the hypercharge related  $g_Y$ in (\ref{axio}) is 
related to the $n_a^2$, $n_d^2$, $n_e^2$ coefficients since the hypercharge depends  on $U(1)^a, \ U(1)^d, \ U(1)^e$. 
Values of the ratios $g_s/f$, $g_Y/f$ constrained by dijet constraints, such that gluonic fusion is achieved where further discsussed in \cite{ibaax}.

\section{Conclusions }

We have shown that the sneutrino can explain the 750 GeV diphoton excess, produced via gluon fusion, within the context of high 
scale D-brane string models. Bottom quarks with charges $1/3 $ contribute to the diphoton excess and mix with the SM down quarks together
with higgsinos. We have not considered the contribution of scalars to the diphoton excess in detail in this work.
 The mass of the 
sneutrino can be naturally as low as 750 GeV, as a result of variation of the complex structure modulus $U_3$ on a 2-dimensional 
torus (turing on a Fayet Iliopoulos term). The variation is small (at least 1 $\%$ of $U_3$) only for values of the string scale $M_s 
\  \in [ 10^{14} \ ,   10^{19} ]$ GeV.
    We have also dicsussed the possibility that the axion is responsible for the diphoton excess. In the present class of models an axion always remain masless as 
it does not get a mass from the generalized Green-Schwarz mechanism involving the Stuelkeberg BF couplings.
In this case, the diphoton signal could be 
produced either from photon fusion or gluon fusion. The couplings of the axions to SU(2) gauge bosons becomes naturally zero, as a result of the 
supersymmetry condition involving the angles between the branes b, c and b, c*, that generates the scalar Higgses needed for electroweak symmetry breaking.
We have also shown that in the case where no coupling between the axions and the colour SU(3) $G^2$ (as considered in \cite{anto3}) exists,  
the zero
 interaction result can be justified in our non-supersymmetric D6-brane model, due to the presence of N=1 supersymmetry in the intersections 
$ab*$, $ac$, $be$, $ce$, where $q_L$, $U_R$, $l_L$, $\nu_R$ are
localized respectively. Thus the superpartners of the $q_L$, $U_R$, $l_L$, $\nu_R$ should also be present in the low 
energy effective action of the 5-stack D6-brane model.

\section{Acknowledgements}
I would like to thank Ignatios Antoniadis for a crucial dicsussion and also Dieter Lust for reading the manuscript.
I would like also to thank Arnold Sommerfeld Center for Theoretical Physics, LMU Munich for its warm hospitality where this work was completed.

\end{document}